\documentclass[11pt]{article}
\input amssym.def
\input amssym.tex

\usepackage{epsfig}

\newcommand{\sect}[1]{\setcounter{equation}{0}\section{#1}}

\def\be{\begin{equation}}
\def\ee{\end{equation}}
\def\ba{\begin{eqnarray}}
\def\ea{\end{eqnarray}}

\begin{document}
\newpage
\bigskip
\hskip 5in\vbox{\baselineskip12pt \hbox{hep-th/0209238}
\hbox{UCTP-105-02}}
\bigskip
\bigskip
\bigskip

\centerline{\bf {\LARGE Higher Dimensional Charged Rotating
Solutions }} \bigskip \centerline{\bf {\LARGE in (A)dS
Space-times}}
\bigskip
\bigskip
\centerline{ \large Adel M. Awad$^{\star}$}
\bigskip
\centerline{\it Department of Physics, University of Cincinnati,
Cincinnati, OH 45221-0011, USA.}
\bigskip
\bigskip
\centerline{$^\star$ email: adel@physics.uc.edu}
\date{September 23, 2002}
\bigskip

\begin{abstract}
We present a class of solutions to the Einstein-Maxwell equations
in $d$--dimensions, all of which are asymptotically (anti)--de
Sitter space-times. They describe electrically charged rotating
solutions, which are generalizations of those found by Lemos
(gr-qc/9404041). These solutions have toroidal, planar or
cylindrical horizons and can be interpreted as black holes, or
black strings/branes. We calculate the inverse temperature and
entropy, and then we use the Brown--York stress--tensor to
calculate mass and angular momenta of these solutions.

\end{abstract}

\sect{Introduction} The AdS/CFT correspondence
\cite{maldacena,witten,gubklebpoly,review}, has created lots of
interest in asymptotically anti--de--Sitter(AAdS) space-times and
their thermodynamics. Several gauge field theory phenomena such as
confinement, confinement/deconfinement phase transitions, and
conformal anomalies, have been shown to be encoded in the
semi--classical physics of AAdS black holes
\cite{witten,witten1,Henningson}. Following these works many AAdS
black configurations have been studied, in the context of the
AdS/CFT correspondence to test this very interesting duality as
well as to understand, more, the physics of the strongly coupled
gauge theory on the boundary (see for example \cite{review}).
Beside the essential role played by AAdS in the above duality,
AAdS enjoys several attractive features. For example, having a
negative cosmological constant $\Lambda<0$ allows for various
types of topologies for black holes horizons in contrast to the
asymptotically flat case with $\Lambda=0$. These horizons can be
spherical, hyperbolic or planar, which by global identifications
can lead to tori, or cylinders (for the planar case) and Riemann
surfaces with genus $g>1$ (for the hyperbolic case), in four
dimensions (see for example \cite{topol}). Another important
property for AAdS is that they are locally thermodynamically
stable, for certain ranges of their parameters. Large
Schwarzschild-AdS black holes--- with horizons
$r_{+}>|\Lambda|^{-1/2}$ --- have positive specific heat because
the AdS effectively acts as a box with reflecting walls provided
by its boundaries \cite{hawkpage}. Another example is the
classical stability of Kerr--AdS black holes, which is not shared
by asymptotically flat Kerr black holes. The reason for their
stability is different from the Schwarzschild case. Kerr--AdS
black holes have rotating Killing vectors which are time--like
everywhere outside the horizon. Following the orbits of these
rotating Killing vectors, thermal radiation can rotate around the
black hole with the same angular velocity of the hole, implying
that super-radiance is not allowed, and the black hole will be in
thermal equilibrium with the thermal gas around it
\cite{hawking2,hawking3}(see also \cite{adel} for other
interesting phenomena in Kerr--AdS space-times).
\\

Since AAdS space--times play such an important role in the above
duality, it is interesting to find new AAdS configurations and
study their semi--classical physics and, if possible, relate it to
the physics of the boundary field theory. In this paper we present
a class of solutions in asymptotically (anti)--de--Sitter
space-times which are electrically charged and rotating. These
solutions are generalizations of the solution found by Lemos
\cite{lemos1,lemos} to higher dimensions. One can interpret these
solutions, depending on their global identifications, either as
black holes with planar horizons in $d$--dimensions, or as black
strings/branes. The paper is organized as follows; first we
present the solutions after writing down the general
Einstein--Maxwell action. Next we discuss the physical properties
of these solutions such as, types of horizons and the extremality
condition. We then calculate the inverse temperature and entropy
of the solutions. Finally we use Brown--York divergence free
stress--tensor to calculate the total mass and angular momenta of
the solutions, after which we give some final remarks.

\section{Rotating Electrically charged $d$--dimension Black Holes}

The action for Einstein-Maxwell theory in $d$--dimensions for
asymptotically (anti)--de--Sitter space-times is

\begin{eqnarray}
I_d= -{1 \over 16 \pi G_d}\int_{\cal M} d^{d}x \sqrt{-g}\left(R+{(d-1)(d-2)
\over l^2}-F^{\mu\nu}F_{\mu\nu}\right),
\end{eqnarray}
where $\Lambda{=}{-(d-1)(d-2)\over 2l^2}$ is the cosmological
constant, $G_{d}$ is Newton's constant in $d$--dimensions, and
$F_{\mu\nu}=\partial_{\mu}A_{\nu}-\partial_{\nu}A_{\mu}$, where
$A_{\mu}$ is the vector potential. Varying the action with respect
to the metric and the vector field $A_{\mu}$ we get the following
field equations \ba
G_{\mu\nu}-\Lambda g_{\mu\nu}&=&2\,T_{\mu\nu},\nonumber\\
\partial_{\mu}\left(\sqrt{-g} F^{\mu\nu}\right)&=&0,
\ea where \be T_{\mu\nu}=F_{\mu \rho}\,F_{\nu}^{\rho}-{1 \over
4}g_{\mu\nu}F^{\mu\nu}F_{\mu\nu}. \ee

Lemos found a four dimensional rotating charged black hole
solution with a flat horizon \cite{lemos1,lemos}, which can be
written in the following form \be
ds^2=-f(r)\,(\Xi\,dt-a\,d\phi)^2+{dr^2 \over f(r)}+{r^2 \over
l^4}(a\,dt-\Xi l^2\,d\phi)^2+{r^2\over l^2} \,d{z}^2 \ee where
$f(r)$ is given by \be
f(r)={r^2 \over l^2}-{m \over r}+{q^2 \over r^2}.\\
\ee The coordinates $\phi$ and $z$ assume the values, $0\leq \phi
< 2\pi $ and $-\infty<z<\infty$. The gauge potential has the form
\be A=-{q \Xi \over r}(dt-{a \over \Xi}d\phi). \ee Here $a$ is the
rotation parameter, $m$ is the mass parameter, $q$ is the electric
charge and $\Xi=\sqrt{1+a^2/l^2}$. This solution has been studied
in \cite{lemos1,lemos,deh,cai,cardoso} and it is known to preserve
some supersymmery \cite{lemoss}. One can observe that there is a
coordinate transformation which relates the rotating solution to
the non--rotating solution, namely, \be t'=\Xi\,t-a\,\phi
\hspace{1.0 in} \phi'={a \over l^2}\,t-\Xi\,\phi \ee This
coordinate transformation is allowed locally but not globally
\cite{lemos1} since it mixes the compactified $\phi$ coordinate
with the time coordinate. As Stachel \cite{stachel} (see also
Lemos \cite{lemos1}) has shown, if the first Betti number of the
manifold is non-vanishing, there are no global diffeomorphisms
that can map the two metrics \cite{lemos1,stachel} and the new
manifold will be parameterized globally by $a$. It is easy to show
that the first Betti number is one for the above solution when we
have cylindrical or toroidal horizons. In higher dimensions we can
have non--vanishing first Betti number by compactifing certain
 numbers of coordinates in a $(d-2)$--dimensional
submanifold of the solution. These are angular coordinates, which
we will call $\phi_i$, describe the rotation of the solutions. The
higher dimensional non--rotating charged solutions, {\it i.e.} $d
> 4$, have been discussed together with their thermodynamics in
\cite{clifford}. In fact, some of the above solutions can be
constructed as near horizon limit to spining D3 or M2 branes
\cite{d2}.
\\

The rotation group in $d$-dimensions is $SO(d-1)$ and the number of independent rotation parameters for a localized object
 is equal to the number of Casimir operators, which is $[(d-1)/2]$, where $[x]$ is the integer part of $x$. We now present our
solution which is the generalization of the above solution found
by Lemos \cite{lemos1,lemos} to arbitrary dimensions with all
rotation parameters. The metric for this solution is given by \ba
ds^2=&&-f(r)\left(\Xi\,dt-\sum_{i=1}^{n}\,a_i\,d\phi_i\right)^2+{r^2\over
l^4}\sum_{i=1}^{n}({a_i }\, dt-\Xi \, l^2\,d\phi_i)^2 +{dr^2 \over
f(r)}\nonumber\\
&&-{r^2 \over l^2}\sum_{i<j}^n
\left(a_i\,d\phi_j-a_j\,d\phi_i\right)^2+{r^2}d\Omega^2 \ea where
$n=[(d-1)/2]$ is the number of rotation parameters $a_i$,
\,$\Xi=\sqrt{1+\sum_i^n \,a_i^2/l^2}$, $d\Omega^2=dy^k dy^k$ is
the Euclidean metric on ($d-2-n$)--dimensions submanifold with
volume $\omega_{d-2}$ and $k=1,...,d-2-n$. Here $f(r)$ is \be
f(r)={r^2\over l^2}-{m \over r^{d-3}}+{q^2 \over r^{2d-6}}, \ee
and the gauge potential is given by \be A=-{q\,\Xi \over
c_d\,\,r^{d-3}}\left(dt-{1 \over \Xi }\sum_i^n a_i\,d\phi_i
\right). \ee The coordinates $\phi_i$ assume the values
$0\leq\phi_i<2\pi$. In the above expression, $c_d$ is a constant
given by \be c_d=\sqrt{{2d-6 \over d-2}}. \ee The asymptotically
de--Sitter version of these solutions can be obtained by simply
taking $l\rightarrow i\,l$ which will be left for future work.

\section{Physical Properties}
\subsection{Horizons and singularities}

Similar to higher dimensional Kerr solutions in asymptotically
flat space-times the above metric has two types of hypersurfaces
or horizons \cite{wald}; a stationary limit surface and an event
horizon. The stationary limit surface is the boundary of the
region in which an observer travelling a long time-like curve can
follow the orbits of the asymptotic time translation Killing
vector $\partial/\partial t$ and so remain stationary with respect
to infinity. Beyond this surface any observer following the orbits
of $\partial/\partial t$ must go faster than the speed of light,
{\it i.e.}, he can not remain stationary with respect to an
observer at infinity. On this surface the Killing vector
$\partial/\partial t$ is null. They are surfaces of infinite
redshift, and are the real solutions of \be
g_{tt}=-\Xi^2\,f(r)+{r^2\over l^2}\sum^{n}_{i=1}\,{a_i^2 \over
l^2}=0 .\ee  On the other hand, surfaces at which the radial
coordinate $r$ flips its signature is an event horizon, and
defined as \be g^{rr}=f(r)=0. \ee If we call the largest solution
of (3.12) $r_{s_+}$, and the outer most event horizon $r_+$. One
can see that for the physically interesting case with $q\,<\, m$,
there is a region between $r_{s_+}$ and $r_+$ in which any
observer following a time-like curve can not remain stationary. To
see this for this solution, let us first simplify the discussion
by keeping only one rotation parameter. Parameterizing the
time-like curve using a proper time $\tau$, we calculate the
norm-square of the tangent $u^a=dx^a/d\tau$ to be negative, {\it
i.e.}, \be g_{tt} (dt/d\tau)^2+2\,
g_{t\phi}(dt/d\tau)(d\phi/d\tau)+g_{\phi\phi}(d\phi/d\tau)^2\,<\,0.\ee
Notice here that in the region between $r_{s_+}$ and $r_+$ all
terms in this expression are positive but the second term, this
can be seen from the fact that $r^2/l^2\,> f(r)$ in this region.
This leads, also, to $g_{t\phi}\,<\,0$ and consequently to
$d\phi/d\tau >\, 0$, {\it i.e.}, any observer is forced to rotate
in the direction rotation of the black hole. These two types of
horizons coincide for static limit of these solutions, {\it i.e.},
as $a \rightarrow0 $, similar to the Kerr solution in
asymptotically flat space-time.

The singularities appeared in the metric by setting $g_{tt}=0$,
and $g^{rr}=0$ are coordinate singularities which can be removed
using more appropriate set of coordinates. One way to check the
existence of physical singularity is to calculate curvature
scalars, such as the square of the Riemann tensor. Calculating
this, we find, \be R^{\mu\nu\rho\sigma}R_{\mu\nu\rho\sigma}\sim
F(r)/r^{(4d-8)},\ee where $F(r)$ is some polynomial in $r$.
Therefore the conclusion is that a physical singularity occurs at
$r=0$.

There is a critical value of the black hole mass parameter, \be
m_{ext}=\left({2d-4\over d-3}\right) \,\,l^{d-1 \over
d-2}\,\left({d-1 \over d-3}\right)^{-{d-1\over
2d-4}}\,\,q^{d-1\over d-2} \ee at which the horizon is degenerate
and the black hole is extremal. We notice here that the extremal
solutions are not the supersymmetric ones. For the solution to be
supersymmetric (in $a=0$ case) we must set $m=0$ \cite{clifford}.

\subsection{Physical Quantities; Temperature Mass and Angular Momenta}

By taking $t\rightarrow i\tau$, we define the Euclidean section of
the metric (2.8) which requires that the period of Euclidean time,
{\it i.e.} the inverse temperature, have the form \be
\beta={4\,\pi\,\Xi\,l^2\, r_{+}^{2d-5} \over
[(d-1)r_{+}^{2d-4}-(d-3)q^2\,l^2]},\ee where $r_{+}$ is the
largest root of $f(r)$. Upon Euclidean continuation, we must take
$a_j\rightarrow i\,a_j$. The identification $\tau \sim \tau+\beta$
leads to the identification $\phi_j \sim \phi_j+i\beta\Omega_j$
\cite{hawking2} as well, where \be \Omega_j ={a_j \over \Xi l^2}
\ee are the angular velocities at the horizon. The entropy for
this metric is \be S={A_{H} \over 4
G_d}={\omega_{d-2}\,r_{+}^{d-2}\,\Xi \over 4 G_d}.\ee To calculate
various thermodynamical quantities we first add the
Gibbons--Hawking boundary term to action. We get Einstein
equations upon varying the bulk metric with fixed boundary metric;
the boundary term is
\begin{eqnarray} I_{\rm s}= -{1 \over 8 \pi G } \int_{\partial
{\cal M}} d^{d-1}x \sqrt{-h} \,K.
\end{eqnarray}
Here $h_{ab}$ is the boundary metric and $K$ is the trace of the
extrinsic curvature $K^{ab}$ of the boundary. We use the
counterterm subtraction method \cite{Henningson,balasubramanian}
to render the action finite, and then use it to calculate
different physical quantities. In this method we add a finite
number of local surface integrals to leave the action finite. For
example, the following counterterm can define a finite action up
to seven dimensions;
\begin{eqnarray} I_{\rm ct}={1 \over 8 \pi
G} \int_{\partial {\cal M}}d^{d-1}x\sqrt{-h}\left[ \frac{(d-2)}{
l}-{l{\cal R} \over 2(d-3)}+... \right]\end{eqnarray} Here ${\cal
R}$ and ${\cal R}_{ab}$ are the Ricci scalar and tensor for the
boundary metric $h$. Having a finite action we can define a
divergence free stress--tensor which has been introduced by Brown
and York \cite{brown} \be T^{ab}={1 \over 8\,\pi G_d}\left[K^{a
b}-h^{ab}K-{(d-2)\over l}h^{ab}+{lG^{ab} \over (d-3)}+...\right].
\ee The above stress--tensor is divergence free for dimensions
less than six, but we can always add more counterterms to have a
finite actions in higher dimensions (see for example
\cite{kraus}). In the case of AAdS solutions with flat horizons,
the only non--vanishing counterterm is the first term in (3.19).
Using the above tress--tensor we can define a conserved charges
for any given AAdS space--time. The conserved charges associated
to a Killing vector $\xi^{a}$ as follows: \be
Q_{\xi}=\int_{\Sigma}d^{d-1}x\sqrt{\sigma}u^{a}T_{ab}\xi^{b},\ee
where $T_{ab}$ is Brown--York stress--tensor, $u_{a}=-Nt,_{a}$,
while $N$ and $\sigma$ are the lapse function and the spacelike
metric which appear in the ADM--like decomposition
 of the boundary metric
 \be ds^2=-N^2\,dt^2+\sigma_{a b}(dx^a+N^a\,dt)(dx^b+N^b\,dt).\ee
 The
 convention we use for the Killing vectors is the following; $\partial_t$ is the
 Killing vector conjugate to $t$ and $\partial_{\phi}$ is the
 Killing vector conjugate to $\phi$.

 Using the above definition for conserved charges, we find the total energy of
 the solution to be given by
 \be
 {\cal M}={\omega_{d-2} \over 16 \pi G_d}[(d-1)\,\Xi^2-1)]\,m,
 \ee
 while the angular momenta are given by
 \be {\cal J}_i={(d-1)\,\omega_{d-2}\over 16 \pi G_d}\,\Xi\,a_i\,m.
\ee We notice here that the total mass reduces to the mass of
charged $d$--dimensions solutions \cite{clifford} upon taking the
limit $a\rightarrow 0$, and for $d=4$ it reduces to the mass and
angular momenta calculated in \cite{lemos,deh}.
\section{Final Remarks}

We have presented a class of charged rotating solutions in
$d$--dimensions which are generalizations to the solutions found
by Lemos. They have toroidal, planar, or cylindrical horizons.
They can also be interpreted as black strings/branes. We have used
the definition of conserved charge provided by the quasilocal
stress--tensor introduced by Brown and York to calculate the total
energy and angular momenta of these solutions. We have also
calculated the inverse temperature and the entropy of these
space--times. It would be interesting to study the thermodynamics
of such solutions and its relevance in the AdS/CFT correspondence.

Another important aspect of these string/branes solutions which
deserves investigation is their classical stability to linearized
perturbations. It known that some extended black configurations
are not classically stable against linearized perturbations
\cite{greg1}, but there are also stable extended configurations
(see for example \cite{kang}). A first step toward a better
understanding of this issue has been provided by Gubser and Mitra
\cite{mitra}; they conjectured that a black string/brane with a
noncompact translational symmetry is classically stable if and
only if it is locally thermodynamically stable. In fact the four
dimensional case given by the metric (2.4) is known to be locally
thermodynamically stable as has been shown in \cite{cardoso,deh},
which leads according to Gubser and Mitra to the classical
stability of the solution. Thus the solutions presented here can
also be used to test the Gubser--Mitra conjecture, as we hope to
discuss in a future work.
\newpage

{\noindent \bf Acknowledgements}\\

We would like to thank Philip Argyres, Paul Esposito, Sumit Das,
Fred Mansouri and Louis Witten for useful discussions, and
especially PA, and PE for suggestions and comments on a
preliminary draft of this paper. This work is supported by funds
provided by the U.S. Department of Energy (D.O.E.) under
cooperative research agreement DE-FG02-84ER-40153.

\end{document}